# Ezhil (எழில்) : A Tamil Programming Language

Muthiah Annamalai, <muthuspost@gmail.com>

**Abstract**

Ezhil(எழில்) is a Tamil language based interpreted procedural programming language. Tamil keywords and grammar are chosen to make the native Tamil speaker write programs in the Ezhil(எழில்) system. Ezhil(எழில்) allows easy representation of computer program closer to the Tamil language logical constructs equivalent to the conditional, branch and loop statements in modern English based programming languages. Ezhil(எழில்)is a compact programming language aimed towards Tamil speaking novice computer users. Grammar for Ezhil(எழில்) and a few example programs are reported here, from the initial proof-of-concept implementation using the Python programming language[1]. To the best of our knowledge, Ezhil language is the first freely available Tamil programming language.

**1. Introduction**

The primary motivation is that like mathematics, computing is a concept, and can be introduced through any native language. Our motivation is toward creating a language to introduce children to computing. Once they know to think in these logical modes (enumeration, recursion, procedural ) then writing any program is an aggregation of ideas in some order.

Ezhil programming system is designed like a procedural language( eg. BASIC / LOGO ), and dynamic typed ( declaration free ) like Ruby/Python; i.e. one can write the program on an interpreter, which is important for students to learn by immediate feedback edit-run-edit-run cycle instead of the edit-compile-re-edit-compile-execute cycle. Ezhil is expected to make computing easy and language semantics follow as closely as possible to reasoning in Tamil by allowing only Tamil keywords, but both Tamil and English identifiers, and function names. In the interest of future availability, the Ezhil programming system is released under free-software under the terms of GNU-GPL including the libraries.

In this paper we discuss the Ezhil language Grammar, Ezhil language keywords, and feature sets in Sec. 3, 4. Details of the Ezhil system implementation are given in Sec. 5, and a brief discussion about missing features and current status of Ezhil language. Ezhil programming language was released for developer testing on Feb. 2008, and freely available since then.

**2. Previous attempts at a Tamil Programming Languages**

Many solutions to the Tamil programming language problem have been suggested earlier. We briefly review the

---

1: This work was done independently of University of Texas at Arlington. The source code for the Ezhil(எழில்) programming language interpreter can be obtained freely through the website [5] or upon request to the author.

proposed solutions using static-typed language, and pre-processor system.

A static-typed Tamil-language system called Swaram [1] was introduced in 2003. Swaram is a full-fledged static-typed programming language, with a feature set resembling C-programming language. Swaram has inspired some of the choice of keywords in the Ezhil language. To its credit Swaram is the first programming language in Tamil, in the true sense with a JIT compiler from source and a virtual machine (VM). In reporting Swaram [3], the authors justify the need for a complete language rather than plain pre-processors, and other syntactic sugar. At time of writing, Swaram is not publicly available, which severely limits language development, system use, community support and improvement. From experience of Python [2] we generally think that interpreted languages are easier for students pickup and enjoy programming with.

Swaram is strongly typed, and allows mixed English & Tamil identifiers. This is a useful feature we have incorporated, as any Tamil programming language system to be useful it needs to provide some form/type of access to allow external libraries which are ( invariably ) written in English. Now this will mean the language has to accept identifiers & functions / variables in a mix of English & Tamil. Keywords can still be Tamil-only. Swaram is the more complete of the previous attempts at providing a Tamil programming language. Ezhil system borrows some of the better ideas freely from the Swaram system.

One such is to use a pre-processors to convert the Tamil-language program into equivalent English language program by a hash-generator for the Unicode (UTF-8) strings into legal strings of another language and compile the resulting program. Pre-processors like Pytham [3] (targets the Python programming language), are not programming languages. They suffer from scaling issues for large projects, demangling symbols, and namespaces among various issues. We need to capture the semantics and the idea of computation through use of meta-syntactic variables. This requires a true programming language with an appropriate builtin functions, and API's.

### 3. Ezhil(எழில்) language Grammar

A part of the grammar for the Ezhil language is presented in a BNF format, and suitable for constructing a recursive descent parser. Most of the terminal symbols are self-explanatory, and are not described here. English equivalents of Tamil keywords are used in the following grammar description.

**Table 1: Ezhil Language Grammar**

| | |
|---|---|
| PROG:  STMT; FUNCDEF<br><br>FUNCDEF : 'def' id '(' ARGLIST ')'  STMTLIST END<br><br>STMTLIST : STMT, STMTLIST \| NONE<br><br>STMT: ASSIGNSTMT<br>ASSIGNSTMT: id := EXPR<br>    copying is a value based deep-copy, not a reference.<br>    later this may change however. | CASE_STMTLIST : 'CASE' '(' EXPR ')' STMTLIST CASE_STMTLIST<br>         \| 'OTHERWISE' STMTLIST<br><br>ARGLIST : ID, ARGLIST \| NONE<br><br>NONE : EOF \| ""<br><br>EXPR : TERM OPTEXPR   \| ARRAY_FULL<br><br>ARRAY_FULL : '[' ARRAY_EXPR* ']'<br>ARRAY_EXPR : expr \| ARRAY_FULL \| NONE |
| STMT: PRINTSTMT<br>PRINTSTMT : 'print' EXPR<br><br>STMT : RETSTMT<br>RETSTMT : 'return' EXPR<br><br>STMT: EVALSTMT<br>EVALSTMT : EXPR<br><br>STMT : WHILESTMT<br>WHILESTMT : 'WHILE' EXPR  STMTLIST 'END'<br><br>STMT : FORSTMT<br>FORSTMT : 'FOR' '(' EXPR1 ',' EXPR2 ',' EXPR3 ')' STMTLIST 'END'<br><br>STMT : SWITCHSTMT<br>SWITCHSTMT : 'SWITCH' '(' EXPR1  ')' CASE_STMTLIST 'END' | OPTEXPR : PLUS_MINUS EXPR  \| e<br><br>ID : REGULAR_ID \| ARRAY_ID<br>ARRAY_ID : REGULAR_ID '[' expr ']'+<br><br>REGULAR_ID : [a-zA-Z]+[0-9]*<br><br>TERM : PRIM OPTTERM<br><br>OPTERM : PROD_DIV TERM   \| e<br><br>PRIM : ( EXPR )  \| NUMBER<br><br>PROD_DIV : * \| /<br><br>PLUS_MINUS : + \| - |

### 3.1 Language features of Ezhil (எழில்)

1. Tamil language keywords for an Interpreted, dynamic programming language.
2. Support recursive functions, and procedural programming. No references are present, and call by value is allowed.
3. Allows mixing english and tamil language identifiers for future compliance with external libraries. Several python builtin functions are accessible directly.
4. Lightweight interpreter implemented on Python, enables extension and usage of various Python based libraries at a later point of time. This is a significant departure from reinventing the wheel as in earlier work, to just using it, by starting on a pre-existing system of Python.
5. Ezhil language is free-software licensed under GNU-GPL, which allows wide distribution and development of the language.

### 4. The Ezhil(எழில்) programming language system

Ezhil(எழில்) keywords and statements are chosen to closely represent in the computer programs, the same chain of reasoning and logic followed in Tamil language itself. The resulting syntax is noticeable feature in that the predicate followed by the expression like in LISP [4], which is a natural way of reasoning by the Tamil language grammar. Conditional Statements are modeled after the IF-ELSEIF-ELSE statement . Variable branch statement statement following SELECT-CASE statement, and several loop control statements deriving from the FOR, WHILE, DO-UNTIL statements are chosen. The function declaration syntax is kept simple. The details are presented in Table. 2. Other statements include  the print statement, and the flow control statements in list (2-4) below.

1. PRINT statement : பதிப்பி
2. BREAK-statement : நேருத்து
3. CONTINUE-statement : தொடர்
4. RETURN-statement : பின்கொடு

**Table 2: Control flow statement Syntax**

| IF-ELSEIF-ELSE statement | SELECT-CASE statement |
|---|---|
| @( X <> Y ) ஆனால்<br>    ##பல விஷயம் செய்க<br>@( அ + ஆ ) இல்லைஆனால்<br>   ##சில செய்க<br>இல்லை<br>   ## கடைசி செய்க<br>முடி | @( அ + ஆ ) தேர்ந்தெடு<br><br>தேர்வு @( ஆ )<br><br>ஏதேனில்<br>   ## கடைசி செய்க<br>முடி |
| FOR Loop | While loop |
| ஆக ( X )<br><br>முடி | @( X < 0 ) வரை<br>   ## பல விஷயம் செய்க<br>முடி |
| Do Until | Function declaration syntax |
| செய்<br>   ##பல விஷயம் செய்க<br>முடியேனில் @( X < 0 ) | நிரல்பாகம் [பெயர்] @( )<br><br>முடி |

Keywords list is summarized in the Table.3 below.

**Table 3: Keyword list**

| | | |
|---|---|---|
| நேருத்து : break<br>ஆக : for<br>ஆனால் : if<br>இல்லைஆனால் : elseif<br>இல்லை : else<br>தொடர் : continue | பின்கொடு : return<br>தேர்ந்தெடு : select<br>தேர்வு : case<br>ஏதேனில் : otherwise<br>வரை: while | செய் : do<br>முடியேனில் : until<br>பதிப்பி : print<br>நிரல்பாகம் : function<br>முடி : end |

Some example programs are given below which work under the Ezhil language system, except the array indexing is not implemented at present.

**Table 4: Simple Ezhil Programs**

| sum all values of x. | Factorial |
|---|---|
| நிரல்பாகம் கூட்டு ( x )<br>  மு = 0; | நிரல்பாகம் fact ( n )<br>  @( n == 0 ) ஆனால்<br>    பின்கொடு 1 |
| ஆக ( டி = 0, டி < len(x), டி = டி+1 )<br>    மு = மு + x[டி]<br>முடி<br><br>பின்கொடு மு<br>முடி | இல்லை<br>    பின்கொடு n*fact( n - 1 )<br>முடி<br>முடி<br>பதிப்பி fact ( 10 ) |
| GCD calculated recursively | Hello program |
| நிரல்பாகம் gcd ( x, y )<br>  மு = max(x,y)<br>  q = min(x,y)<br>  @( q == 0 ) ஆனால்<br>    பின்கொடு மு<br>  முடி<br>  பின்கொடு gcd(மு-q,q)<br>முடி<br><br>gcd( 10, 15) | "வணக்கம்!"<br>பதிப்பி "நக்கீரண் அழைக்கிரது" |

## 5. Ezhil System Implementation

At the time of writing Ezhil language is implemented in a object oriented fashion in the Python programming language. The core of the interpreter is based on a handwritten lexical analyzer and recursive descent parser with various objects of the abstract syntax tree implementing a visitor pattern for easy evaluation. Ezhil converts the program code into Python AST objects and executes them as Python objects with full access to the Python programming language environment.

The total size of the code is around 1350 lines of code without comments. The interpreter code is licensed under GNU-GPL, and can be downloaded for development purposes from URL [5]. Ezhil system allows access to Python builtin functions (basic math, string functions), and possible extensions to the rest of Python libraries in similar manner.

Future work is to add language support for Arrays, variable arguments, lists, dictionaries, to provide support for general purpose storage. Array types need strong type checking, matching indices, and initializing the types which means the language can be more general once we have array support. Documentation for the standard library & builtin definitions borrowed from Python is missing right now. On similar lines we also need to provide supporting documentation, user guides, and a language reference manual. A Native Tamil character library to read, sort, write stream processing directly in Tamil letters, would be very useful. Using this we could recognize word endings, beginnings etc for Tamil text processing.

## 6. Conclusion

Ezhil is a promising candidate for a general purpose interpreted programming language in Tamil. Ezhil is currently a work in progress with several new and consistent features. We have reported the successful prototyping of Ezhil in Python.